\documentclass[a4paper,onecolumn,american]{revtex4}
\usepackage[T1]{fontenc}
\usepackage[utf8]{inputenc}
\usepackage{amsmath}
\usepackage{amssymb}
\usepackage{graphicx}
\usepackage{esint}
\def\p{\ensuremath{\mathbb P}}
\def\N{\ensuremath{\mathbb N}}

\makeatletter

\pdfpageheight\paperheight
\pdfpagewidth\paperwidth

\makeatother

\usepackage{babel}
\begin{document}

\title{Mixing properties in the advection of passive tracers via recurrences
and extreme value theory}

\author{Davide Faranda}

\affiliation{Laboratoire SPHYNX, Service de Physique de l'Etat Condens\'e, DSM,
CEA Saclay, CNRS URA 2464, 91191 Gif-sur-Yvette, France}

\email{davide.faranda@cea.fr}

\author{Xavier Leoncini}

\affiliation{Aix Marseille Universit\'e, CNRS, CPT, UMR 7332, 13288 Marseille, France }

\affiliation{Universit\'e de Toulon, CNRS, CPT, UMR 7332, 83957 La Garde, France }

\email{Xavier.Leoncini@cpt.univ-mrs.fr}

\author{Sandro Vaienti}

\affiliation{Aix Marseille Universit\'e, CNRS, CPT, UMR 7332, 13288 Marseille, France }

\affiliation{Universit\'e de Toulon, CNRS, CPT, UMR 7332, 83957 La Garde, France }

\email{Sandro.Vaienti@cpt.univ-mrs.fr}

\begin{abstract}
In this paper we characterize the mixing properties in the advection  of   passive tracers by exploiting the extreme value theory for dynamical systems. With respect to classical techniques directly related to the Poincar\'e recurrences analysis, our method provides reliable estimations of the characteristic mixing times and distinguishes between barriers and unstable fixed points. The method is based on a check of convergence for extreme value laws on finite datasets. We define the mixing times in terms of the shortest time intervals such that extremes   converge to the asymptotic (known) parameters of the Generalized Extreme Value distribution model. Our technique is suitable for applications in the analysis of other systems where mixing time scales need to be determined and limited datasets are available.
\end{abstract}
\maketitle

\section{Introduction}

A general purpose of dynamical systems theory is to characterize the
stability properties of orbits. The distinction between regular and
chaotic dynamics can be easily made in a dissipative case whereas
for conservative systems it is usually an hard task especially when
we are in presence of many degrees of freedom or for a complex geometry
of the phase space. A large number of tools known as indicators of
stability have been developed for this purpose; Lyapunov Characteristic
Exponents (LCEs) \citet{wolf1985determining}, \citet{rosenstein1993practical},
\citet{skokos2010lyapunov} and the indicators related to the Return
Time Statistics \citet{kac1934notion}, \citet{gao1999recurrence},
\citet{hu2004statistics}, \citet{buric2005statistics} have been
used for a long time for such a task. Nevertheless, in the recent
past, the need for computing stability properties with faster algorithms
and for systems with many degrees of freedom resulted in a renewed
interest in the technique and different dynamical indicators have
been introduced. The Smaller Alignment Index (SALI) described in \citet{skokos2002smaller}
and \citet{skokos2004detecting}, the Generalized Alignment Index
(GALI), introduced in \citet{skokos2007geometrical} and the Mean
Exponential Growth factor of Nearby Orbits (MEGNO) discussed in \citet{cincotta2003phase},
\citet{gozdziewski2001global} are suitable to analyze the properties
of a single orbit. They are based on the divergence of nearby trajectories
and require in principle the knowledge of the exact dynamics. Another
class of indicators is based on the round off error properties and
has been discussed in \citet{faranda2011reversibility}: the divergence
between two trajectories starting from the same initial condition
but computed with different numeric precision can be used to illustrate
the dynamical structure. The so called Reversibility Error that measures
the distance between a certain initial condition and the end point
of a trajectory iterated forward and backward for the same number
of time steps give basically the same information.

These indicators perform generally well when chaotic and regular trajectories
are  separated.  However, in some interesting
physical problems, a key question is to distinguish between different
mixing regions and possibly recognize the associated mixing time-scales.
This problem becomes  extremely relevant for describing
the dynamics of a passive tracer in an array of alternating vortices.
This type of flows emerge from Rayleigh-B\'enard convection and can
be studied as well experimentally using electromagnetic forces~\citep{solomon1988chaotic,solomon2003lagrangian,willaime1993spatiotemporal}.
One of the primary interest in this type of flow is that it can be
generated by different instabilities, and as such it lays on one of
the paths to turbulence. When considering mixing or transport in fluids,
one usually rely more on a Lagrangian than a Euler point of view.
As such regarding transport, the dynamics of passive tracers advected
in a two dimensional incompressible flow is Hamiltonian. In this setting
the canonical variables are directly the space ones, allowing for
a direct visualization of the phase space. Moreover we have an accessible framework
to test theoretical ideas and have a direct grasp of their physical
consequences.\\
 In our study we consider the flows proposed in \citet{benzekri2006chaotic,bachelard2007targeted}.
These flows theoretically offer the peculiarity of ``targeted mixing'',
meaning that mixing is efficiently performed within cells formed by
one dimensional barriers. This offers a suitable setting to test our
approach and quantify it versus the already proposed measurements.
Before moving one, we give  a brief description of the flows.
We first start with the  integrable stream function that describes
an array of alternating vortices .
\begin{equation}
\Psi_{0}(x,y)=\alpha\sin x\sin y.\label{psi0}
\end{equation}
Here  the $x$-direction is the
one along the channel and the $y$-direction is the bounded orthogonal
one. The amplitude $\alpha$
sets the maximal value the velocity. The
dynamics resulting from the Hamiltonian which is identical to the
stream function Eq.~(\ref{psi0}) is integrable and thus no chaotic
mixing occurs. In this setting, tracers motion are confined within
barriers delimited by the invariant lines $y=\pi$ and $y=0$ and
the other invariant lines which are localized at $x=m\pi$ for $m\in\mathbb{Z}$
along the channel. As we shall see later, it is important to mention
that these points are joined by vertical heteroclinic connections
for which the stable and unstable manifolds coincide. The phase space, which is here the real space,
is then characterized by a channel populated by alternating vortices
with separatrices localized at $x=m\pi$ for $m\in\mathbb{Z}$. 

In order to visualize the phenomenon of chaotic advection
in experiments~\citep{solomon1988chaotic,solomon2003lagrangian,willaime1993spatiotemporal},
a typical perturbation $f(x,y,t)$ is introduced as a time dependent
forcing. This allows for instance to subsequently study the transport
and mixing properties. To be more explicit, the perturbation corresponds
to a modification of the stream function~: 
\begin{equation}
\Psi_{c}(x,y,t)=\Psi_{0}(x+f(x,y,t),y).\label{psic}
\end{equation}
For instance, in the experiment a flow has been realized which is
well modeled by the following stream function~\citep{solomon1988chaotic}~:
\begin{equation}
\Psi_{1}(x,y,t)=\alpha\sin(x+\epsilon\sin\omega_{0}t)\sin y.\label{courant}
\end{equation}
The perturbation $f$ becomes simply $f=\epsilon\sin\omega_{0}t$
and describes the lateral oscillations of the roll patterns where
$\epsilon$ and $\omega_{0}$ are respectively the amplitude and the
angular frequency of the lateral oscillations. Setting the proper
time units, we may assume that $\omega_{0}=1$. In this setting, the field lines have not changed
but simply oscillate back and forth along the channel, and chaotic
advection is triggered. This perturbation breaks the separatrices
and invariant tori, leading to chaotic mixing along the channel. However
we still have invariant tori  corresponding to stable island of regular motion
near the vortex cores. These act as transport barrier and mixing is
not uniform. Regarding this problem a different perturbation was proposed
in \citet{bachelard2007targeted,benzekri2006chaotic}. This perturbation
allows in some windows of parameters to only preserve the separatrices
while destroying all regular tori. The separatrices acting as transport
barrier we end up with a homogeneous mixing within cells delimited
by the separatrices. The proposed perturbation writes~: 
\begin{equation}
\Psi_{c}(x,y,t)=\alpha\sin[x+\epsilon\sin t+\alpha\cos yC_{\epsilon}(t)]\sin y,\label{Strmcont}
\end{equation}
where 
\begin{equation}
C_{\epsilon}(t)=\sum_{n\geq0}\frac{-2}{2n+1}\mathcal{J}_{2n+1}(\epsilon)\cos(2n+1)t,\label{serbessl}
\end{equation}
and $\mathcal{J}_{l}$ (for $l\in\mathbb{N}$) are Bessel functions
of the first kind. 

We defer to \citet{bachelard2007targeted} for the
details but we recall that this perturbation has two main purposes:
i) particles remain trapped within a specific domain bounded by two oscillating
barriers (suppression of chaotic transport along the channel), ii)
the stochastic sea seems to cover the whole domain (enhancement of
mixing within the cells).\\

 In \citet{bachelard2007targeted,benzekri2006chaotic},
the mixing properties of the barriers have been analyzed for several
specific values of $\alpha,\epsilon$. For instance the finite time
Lyapunov map was computed, showing some kind of uniform mixing. When
performing an analysis using Poincar\'e recurrences and a finite time
average recurrence time, the barriers naturally emerged however no
remarkable differences between the barriers and the fixed points stood
out.

In this paper we try to overcome this difficulty by suggesting a new method for the characterization of mixing times based on the Extreme Value Theory
(EVT) for dynamical systems. In particular, we will show that by exploiting the EVT, one is able to observe   differences between
unstable fixed points and barriers as well as to extrapolate the characteristic
time scales such that the dynamics become mixing. 

The EVT was originally introduced by \citet{fisher},
\citet{gnedenko} to study the maxima of a series of independent and
identical distributed variables: under very general hypothesis a limiting
distribution called Generalized Extreme Value (GEV) distribution exists
for the series of extremes. An extensive account of recent results
and relevant applications is given in \citet{ghil2010extreme}. 

In the recent past the EVT has been adapted to study the output of
dynamical systems. As we will explain in detail in the next section,
it is not trivial to observe   asymptotic GEV distributions in dynamical
systems: some sort of independence of maxima must be recovered by
requiring certain mixing conditions on the orbits. Furthermore, we
need to introduce some peculiar observables that satisfy the condition
proposed by \citet{gnedenko} on the parent distribution of data: they are
related to the closest return of a trajectory in a ball centered around
a reference point of the attractor and therefore allow a very detailed
description of the dynamics in the neighborhood of the chosen point.
 
The parameters of the distribution are dependent on the geometrical
properties of the system i.e. the local dimension of the attractor
\citet{freitas}, \citet{lucarini2012extreme}. When this properties
are known, like for the advection of the passive tracers, one can
study the convergence of finite datasets to the asymptotic parameters recovered in the limit of infinitely long time series.  For finite datasets, the rate of convergence will be directly related to the  chosen point of the phase space and to the  local mixing structure of the trajectories passing nearby.  The main idea is to define a mixing time scale based on  the minimum time interval for the selection of maxima  which allows for recovering the asymptotic parameters predicted by the theory.  We will tackle this problem in the remaining of the paper which is organized
as follows: in Section 2 we recall the results of EVT for dynamical
systems and explain the numerical algorithm and procedures used to
compute the parameters of the GEV distribution introduced in \citet{faranda2012generalized}.
In Section 3 we describe the model and the numerical results. Section
4 is dedicated to discussion and possible outlooks

\section{Asymptotic Extreme Value Theory for dynamical systems}

 \citet{gnedenko} studied the convergence of maxima of i.i.d.
variables 
\[
X_{0},X_{1},..X_{m-1}
\]
with cumulative distribution function (cdf) $F(x)$ of the form: 
\begin{equation}
F(x)=P\{a_{m}(M_{m}-b_{m})\leq x\} \label{cum}
\end{equation}
where $a_{m}$ and $b_{m}$ are normalizing sequences and $M_{m}=\max\{X_{0},X_{1},...,X_{m-1}\}$.
Eq.~\ref{cum} can be rewritten as $F(u_{m})=P\{M_{m}\leq u_{m}\}$
where $u_{m}=x/a_{m}+b_{m}$. Under general hypothesis on the nature
of the parent distribution of data, the cdf of maxima $F(x)$ converges
to a single family of generalized distribution called GEV distribution
with cdf:

\begin{equation}
F_{G}(x;\mu,\sigma,\kappa)=\exp\left\{ -\left[1+{\kappa}\left(\frac{x-\mu}{\sigma}\right)\right]^{-1/{\kappa}}\right\} ;\label{cumul}
\end{equation}

which holds for $1+{\kappa}(x-\mu)/\sigma>0$, using $\mu\in\mathbb{R}$
(location parameter) and $\sigma>0$ (scale parameter) as scaling
constants in place of $b_{m}$, and $a_{m}$ \citep{pickands}. In
particular, in \citet{faranda2011numerical} we have shown that the
following relations hold:

\[
\mu=b_{m}\qquad\sigma=\frac{1}{a_{m}}.
\]

${\kappa}\in\mathbb{R}$ is the shape parameter also called the tail
index and discriminate the type of classical extreme value laws:
when ${\kappa}\to0$, the distribution corresponds to a Gumbel type
( Type 1 distribution). When ${\kappa}$ is positive, it corresponds
to a Fr\'echet (Type 2 distribution); when ${\kappa}$ is negative, the extreme value law
corresponds to a Weibull (Type 3 distribution).\\

In the last decade many works focused on the possibility of treating
time series of observables of deterministic dynamical system using
EVT. The first rigorous mathematical approach to extreme value theory
in dynamical systems goes back to the pioneer paper by  \citet{collet2001statistics}.
Important contributions have successively been given in \citet{freitas2008},
\citet{freitas}, \citet{freitas2012extremal} and by \citet{gupta2011extreme}.
The goal of all these investigations was to associate to the stationary
stochastic process given by the dynamical system, a new stationary
independent sequence: when the latter sequence satisfies one of the
classical three extreme value laws, the
same result also holds for the original dynamical sequence. We summarise shortly the main findings of the theory.\\

Let us consider a dynamical systems $(\Omega,{\cal B},\nu,f)$, where
$\Omega$ is the invariant set in some manifold, usually $\mathbb{R}^{d}$,
${\cal B}$ is the Borel $\sigma$-algebra, $f:\Omega\rightarrow\Omega$
is a measurable map and $\nu$ a probability $f$-invariant Borel
measure.\\
 In order to adapt the extreme value theory to dynamical systems,
we introduce the stationary stochastic process $X_{0},X_{1},...$ given
by:

\begin{equation}
X_{m}(x)=g(\mbox{dist}(f^{m}(x),\zeta))\qquad\forall m\in\mathbb{N},\label{sss}
\end{equation}

where 'dist' is a distance on the ambient space $\Omega$, $\zeta$
is a given point and $g$ is an observable function. The probability
measure is here the relevant invariant measure $\nu$ for the dynamical
system often called the physical measure. Hereinafter we will
use three types of observables $g_{i},i=1,2,3$ that are suitable
to obtain one of the three types of extreme value laws for normalized
maxima:

\begin{equation}
g_{1}(x)=-\log(\mbox{dist}(x,\zeta)),\label{g1}
\end{equation}

\begin{equation}
g_{2}(x)=\mbox{dist}(x,\zeta)^{-1/\beta},\label{g2}
\end{equation}

\begin{equation}
g_{3}(x)=C-\mbox{dist}(x,\zeta)^{1/\beta},\label{g3}
\end{equation}

where $C$ is a constant and $\alpha>0\in\mathbb{R}$ \citet{collet2001statistics},
\citet{freitas}.\\

By using these observables we get convergence to the Type 1, 2, or 3
distribution if one can prove two sufficient conditions called $D_{2}$
and $D'$ which basically require a sort of independence of the stochastic
dynamical sequence in terms of uniform mixing conditions on the distribution
functions. In particular condition $D_{2}$, introduced in its actual
form in \citet{freitas2008}, could be checked directly by estimating
the rate of decay of correlations for a suitable class of observables. We summarize these conditions as follows:\\

 If $X_{m},m\ge0$ is our stochastic process, we can define $M_{j,l}\equiv\max\{X_{j},X_{j+1},\cdots,X_{j+l}\}$
and set $M_{0,m}=M_{m}$. The condition $D_{2}(u_{m})$ holds for
the sequence $X_{m}$ if for any integer $l,t,m$ we have 
\[
|\nu(X_{0}>u_{m},M_{t,l}\le u_{m})-\nu(X_{0}>u_{m})\nu(M_{t,l}\le u_{m})|\le\gamma(m,t),
\]
where $\gamma(m,t)$ is non-increasing in $t$ for each $m$ and $m\gamma(m,t_{m})\rightarrow0$
as $m\rightarrow\infty$ for some sequence $t_{m}=o(m)$, $t_{m}\rightarrow\infty$.\\
Let $(k_n)_{n\in\N}$ be a sequence of integers such that
\begin{equation}
\label{eq:kn-sequence-1}
k_n\to\infty\quad \mbox{and}\quad  k_n t_n = o(n).
\end{equation}
We say that $D'(u_n)$ if there exists a sequence $\{k_n\}_{n\in\N}$ satisfying \eqref{eq:kn-sequence-1} and such that
\begin{equation}
\label{eq:D'un}
\lim_{n\rightarrow\infty}\,n\sum_{j=1}^{\lfloor n/k_n \rfloor}\p( X_0>u_n,X_j>u_n)=0.
\end{equation}

Here $\lfloor m/l\rfloor$ indicates the integer part of $m/l$.\\
 Instead of checking the previous conditions, we can use other results
that established a connection between the extreme value laws and the
statistics of first return and the Hitting time statistics (hereinafter HTS)   \citep{freitas,freitasNuovo}. Before introduing the HTS, we need first to define
the recurrence time $\tau_{A}$ in a measurable set $A\in\Omega$,
as

\[
\tau_{A}(x)=\inf_{t\geq1}\left\{ x\in A:f^{t}(x)\in A\right\} ,
\]

and the average recurrence time $<\tau_{A}>$ as

\[
<\tau_{A}>=\int\tau_{A}(x)\mbox{d}\mu_{A}(x)\qquad\mu_{A}(B)=\frac{\mu(A\cap B)}{\mu(A)}.
\]

We notice that, whenever the measure $\mu$ is ergodic, Kac'theorems
ensures that $<\tau_{A}>=\mu(A)^{-1}.$ Following \citet{hiratavaienti}
and \citet{buric2003weak}, we define the HTS as the following limit
(whenever it exists):

\begin{equation}
H(t)=\lim_{\mu(A)\to0}\mu_{A}(A_{>t})\qquad A_{>t}\equiv\left\{ x\in A:\frac{\tau_{A}(x)}{<\tau_{A}>}>t\right\} .\label{RTS}
\end{equation}

In particular, \citet{freitas} and \citet{freitasNuovo} showed that
for dynamical systems preserving an absolutely continuous invariant
measure or a singular continuous invariant measure $\nu$, the existence
of an exponential HTS on balls around almost any
point $\zeta$, namely $H(t)=e^{-t},$ implies the existence of extreme
value laws for one of the observables of type $g_{i},i=1,2,3$ described
above. The converse is also true, namely if we have an extreme value
law which applies to the observables of type $g_{i},i=1,2,3$ achieving
a maximum at $\zeta$, then we have exponential HTS
to balls with center $\zeta$. Recently, these results have been generalized
to local returns around balls centered at periodic points \citet{freitas2012extremal}
and for stochastically perturbed dynamical systems \citep{faranda2013extreme,faranda2013extremenon,faranda2013recurrence}.\\

\section{The method for finite time datasets}

In \citet{faranda2011numerical} and \citet{lucarini2012extreme},
the authors have analyzed both from an analytical and a numerical point
of view the Extreme Value distribution in a wide class of low dimensional
maps showing that, when the conditions $D'$ and $D_{2}$ are verified,
the block maxima approach can be used to study extrema. This approach consists
of dividing the data series of length $k$ of some observable into
$n$ bins each containing the same number $m$ of observations, and
selecting the maximum (or the minimum) value in each of them \citep{coles}. The GEV distribution is obtained by performing a fit of the histogram of maxima (minima) to the GEV model. When one uses the $g_{i}$ observable functions and the underlying dynamic is mixing, the asymptotic GEV parameters are known and depend on $m$
(or equivalently $n$) and the local dimension of the attractor $d$. In particular, the following equations hold:

For $g_{1}$ type observable:

\begin{equation}
\sigma=\frac{1}{d}\qquad\mu\sim\frac{1}{d}\ln(k/n)\qquad\kappa=0\label{g1res}.
\end{equation}

For $g_{2}$ type observable:

\begin{equation}
\sigma\sim n^{-1/(\alpha d)}\qquad\mu\sim n^{-1/(\alpha d)}\qquad\kappa=\frac{1}{\beta d}\label{g2res}.
\end{equation}

For $g_{3}$ type observable:

\begin{equation}
\sigma\sim n^{1/(\alpha d)}\qquad\mu=C\qquad\kappa=-\frac{1}{\beta d}\label{g3res}.
\end{equation}

Here $\sim$ means asymptotically for $m,n \to \infty$. At finite time, the convergence depends on the rate of mixing around the point $\zeta$. We can have one of the following behavior:
\begin{itemize}
\item For $\zeta$s on periodic or quasi-periodic orbits we do not observe convergence
to the GEV distribution. If the motion is purely periodic, the asymptotic extreme value law is a Dirac's
delta otherwise it is a collection of Heaveside functions modulated by the shape of the $g_{i}$s.
\item For $\zeta$s on mixing orbits  there exists a value of $m$ such that the previous Eqs.~\ref{g1res}-\ref{g3res} hold. This value of $m$ can be defined as the shortest mixing time scale. As we will see from the numerical analysis, detailed mixing properties of the phase space  may be explored with this method.
\item For $\zeta$s  located in the proximity of  unstable fixed points, the previous Eqs~\ref{g1res}-\ref{g3res} do not hold.  Extremes cluster around the fixed points making the convergence as slower as closer we get to the fixed point. 
\end{itemize}

\subsection{A practical numerical algorithm}

In \citet{faranda2012generalized} we have introduced a simple algorithm
to get the parameter specified in Eqs~\ref{g1res}-\ref{g3res}:
\begin{enumerate}
\item Compute the orbit of the dynamical system for $k$ iterations. 
\item Compute the series $X_{m}(x)=g(\mbox{dist}(f^{m}(x),\zeta))$ where
$\zeta$ is a point of the phase space. 
\item Divide the series in $n$ bins each containing $m$ data. 
\item Take the maximum in each bin and fit the GEV distribution. 
\end{enumerate}
For the inference, we have used the Maximum Likelihood Estimation (MLE)
procedure explained in \citet{faranda2012generalized}. There the authors introduced the method and tested it on the relevant
example of the Standard map. They characterized  different 
regions of the phase space in terms of  rate of  convergence  to the parameters expected by Eqs.
\ref{g1res}- \ref{g3res}. The results have been checked against
the Divergence of two nearby trajectories and the Reversibility error
as introduced in \citet{faranda2011reversibility}.  Once the experimental parameters $\mu,\ \sigma,\ \kappa$ are obtained by a fit at a certain $m$, there are only two possibilities: 
\begin{itemize}
\item If the fit succeeds one can repeat the experiment for shorter bin
lengths and find the smallest $m$ such that, for the chosen $\zeta$,
the fit converges. This defines the shortest mixing time scale. 
\item If the fit fails one should repeat the experiment by increasing the
size of $m$ until it is possible to retain a sufficient number of
maxima to perform a reliable fit to the GEV model. 
\end{itemize}
As we have already said, for purely periodic orbit one never finds a $m$ such that the fit
converges.  In the next section we show how the application of this method provides reliable results in the case of the advection of a passive tracer.

\section{Results}

We present the results obtained for the Hamiltonian dynamics associated
to the stream function in Eq.~\ref{Strmcont}. By definition, the advection term
 relates to the action of being moved by and with a
flow. The velocity field is then obtained by ${\bf v}=\mathrm{curl}(\psi\;\hat{{\bf z}})$,
where $\hat{{\bf z}}$ is the unit vector normal to the flow. The
flow of passive tracers exhibits a Hamiltonian structure~:
\begin{equation}
\dot{x}=-\frac{\partial\Psi}{\partial y},\hspace{1.2cm}\dot{y}=\frac{\partial\Psi}{\partial x}\:,\label{eq:Hamilton_advec}
\end{equation}
where $(x,y)$ corresponds to the coordinates of the tracer on the
plane. The space variables $(x,y)$ are canonically conjugate for
the stream function $\Psi$ which acts as the Hamiltonian of the system.
Hence the phase space is formally the two dimensional physical space
(with the addition of time).\\

\begin{figure}
\includegraphics[width=18cm]{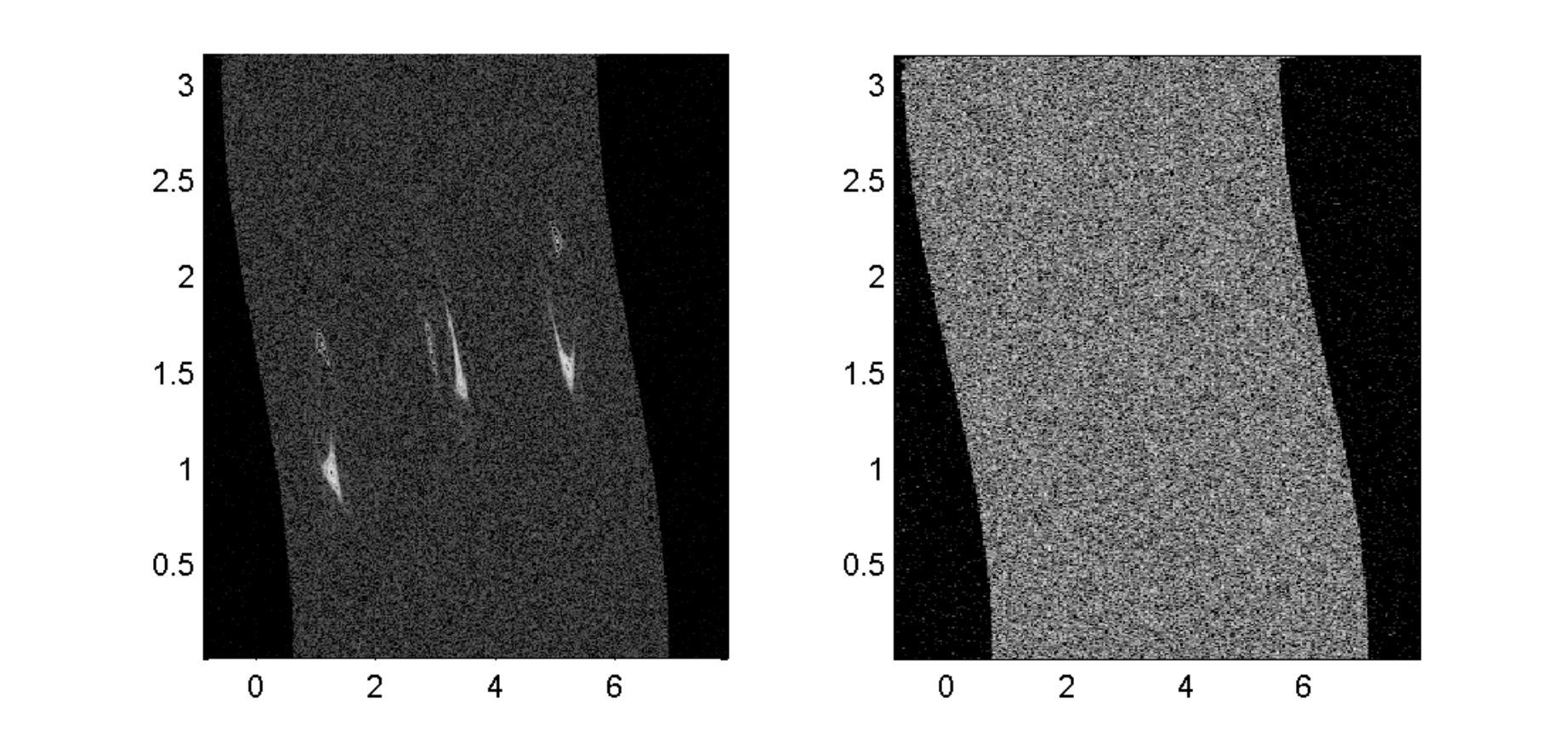} \caption{ Poincar\'e sections for the stream function in Eq.~\ref{Strmcont}.
The parameters are $\alpha=1,\epsilon=0.63$ (left) and $alpha=1,\epsilon=0.8$
(right). }

\label{poincare} 
\end{figure}

In Fig.~\ref{poincare} two Poincar\'e section obtained by numerical
In Fig.~\ref{poincare} two Poincar\'e sections obtained by numerical
have been performed by setting the time step $\Delta T=5\cdot10^{-3}$,
and computing the trajectory of 1000 particles, released at $x_{0}=3.3,\ y_{0}=1.6$
for 1000 time iterations. The left panel of Fig \ref{poincare} refers
to the set of parameters $\alpha=1,\epsilon=0.63$, the right one
to $\alpha=1,\omega=0.8$. For the first set of parameters, stability
islands are clearly visible in the domain whereas for $\alpha=1,\omega=0.8$,
the domain looks well mixed on the time scale considered.

\begin{figure}
\includegraphics[width=18cm]{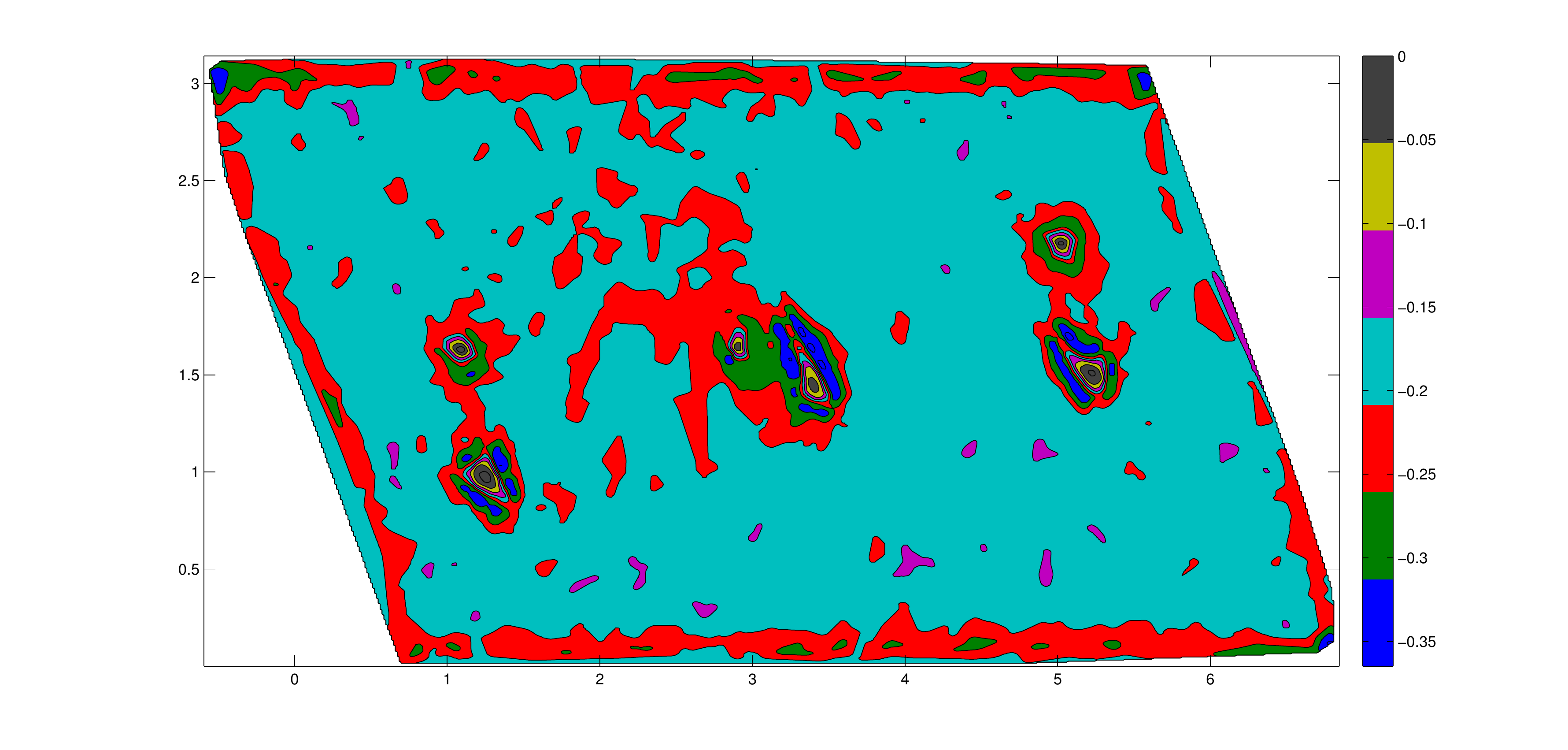} \caption{$\kappa$ for the observable $g_{3}$, $\beta=3$ for an ensemble
of 2000 points. The theoretical expected value is $\kappa=-1/6$ (azure).
See text for a description.}

\label{dyn} 
\end{figure}

For the extreme value analysis we consider the experiment with $\epsilon=0.63$.
First one can check that the method based on the EVT allows for recognizing
the different stability regions. The experiment follows the set up
described before: we consider a very long run - $s=10^{6}$ iterations
- and take 2000 different $\zeta$ points uniformly distributed in the domain at which the extreme value
statistics for the observables $g_{i}$ is computed by taking $m=250,n=4000$
maxima. The results obtained with the EVT analysis are qualitatively
similar for the three observables considered so that in Fig. \ref{dyn}
we have chosen to represent only the results for the observable $g_{3}$
with $\beta=3$. Since we are dealing with a bi-dimensional system,
we expect to find $\xi=-1/6$ in the chaotic region. This is true
in a wide region of the phase space which can be identified as a region where orbits mix efficiently. However we observe a different behavior in correspondence of the regular islands where divergence from
the theoretical expected parameters are observed. In Fig. \ref{dyn} the locations
of the barriers is well highlighted as it forms a sort of frame of divergent values of $\kappa$ around the figure (red areas). Within the barriers the values of the shape parameters
are slightly different than in the chaotic sea and points to regions
where intermediate properties between the regular islands and the
chaotic sea are present.

\begin{figure}
\includegraphics[width=18cm]{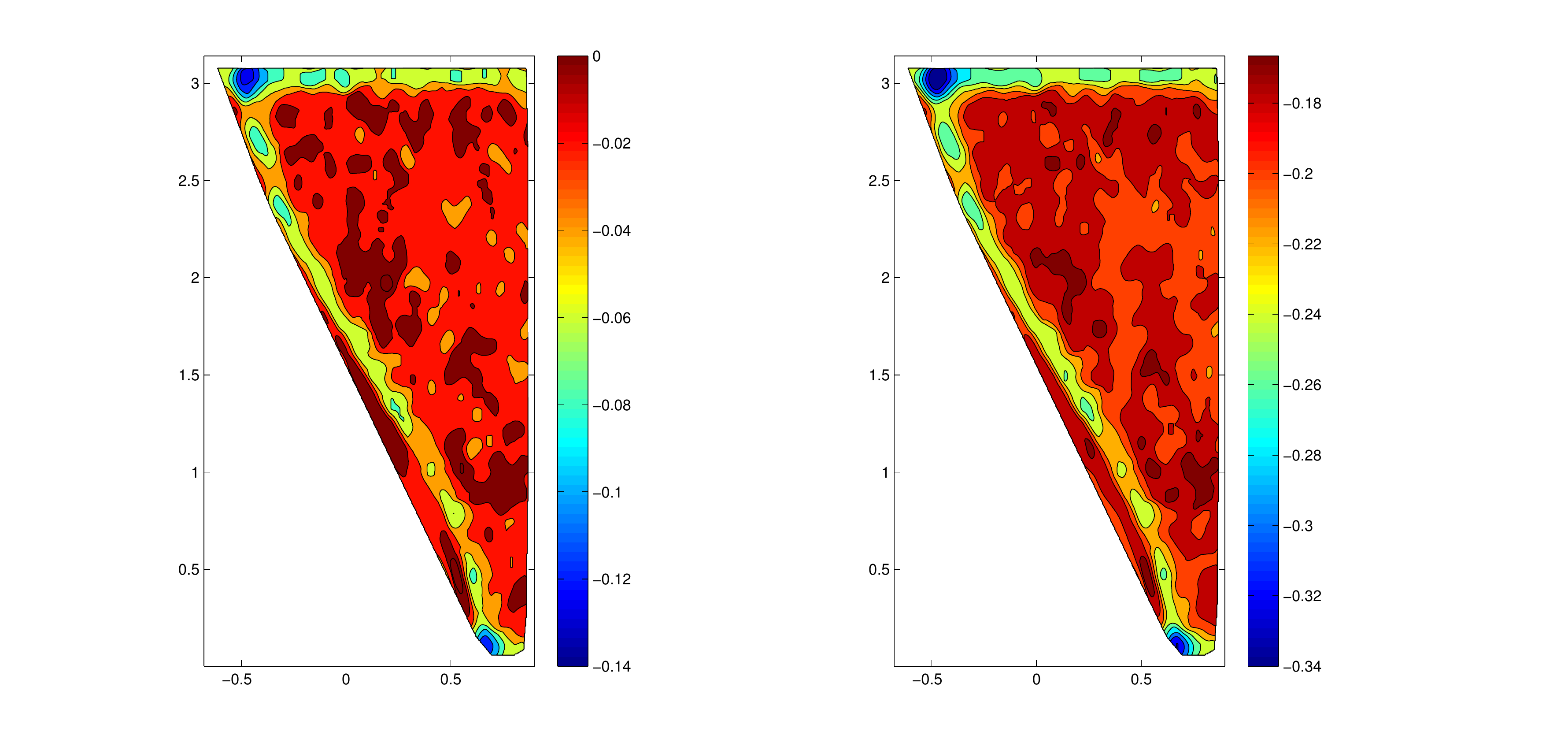} \caption{ Left: $\kappa$ for the observable $g_{1}$. Right: $\kappa$ for
the observable $g_{3}$, $\beta=3$. Ensemble of 2000 points. The
theoretical expected value are $\kappa=0$ for $g_{1}$ and $\kappa=-1/6$
for $g_{3}$. See text for a description.}

\label{bar} 
\end{figure}

In order to better quantify the effect of the barriers, we have isolated
a region of the phase space corresponding to the left border of the
plot and repeated the analysis. The results are presented in Fig.
\ref{bar} for $g_{1}$ (left) and $g_{3}$ (right). The expected
values in the chaotic see are respectively $0$ and $-1/6$ and they
correspond to the highest values of the color-bars. The position of
the barriers and the in-homogeneities within them are evident. In
particular, one can recognize the fixed points (top left) and (bottom
right) for the remarkable deviations from the theoretical parameters.
Other methods (i.e. the Poincar\'e recurrences technique applied in
\citet{bachelard2007targeted}) do not distinguish between barriers
and fixed points; actually the latter method quantify the frequency
of visits in small balls spread in the phase space. Instead the parameters
of the EVT depend on the structure of the fixed points via the so
called extremal index $\theta$, introduced by \citet{freitas2012extremal}
and further analyzed by \citet{faranda2013extremenon}. By invoking
again the equivalence between extreme value statistics and statistics
of return times, the presence of an extremal index simply means that
the statistics of the first return time in a ball shrinking to zero
around a periodic point of (minimal) period $p$ and normalized with
the average recurrence time, converge to $e^{-t\theta}$, where $\theta:=\phi(p)$
is a non-linear function $\phi$ of the period $p$ determined by
the potential associated to the invariant measure $\mu$ \cite{haydn2009compound}. Instead the distribution
is simply the exponential one $e^{-t}$ around a non-periodic point discussed after Eq.~\ref{RTS}.

This divergence between the behavior around   periodic and non-periodic
points reflects in a departure of the theoretical parameters expected
for the GEV in the regime of pure Gumbel's law shown in  Fig.~\ref{bar}.

\begin{figure}
\includegraphics[width=18cm]{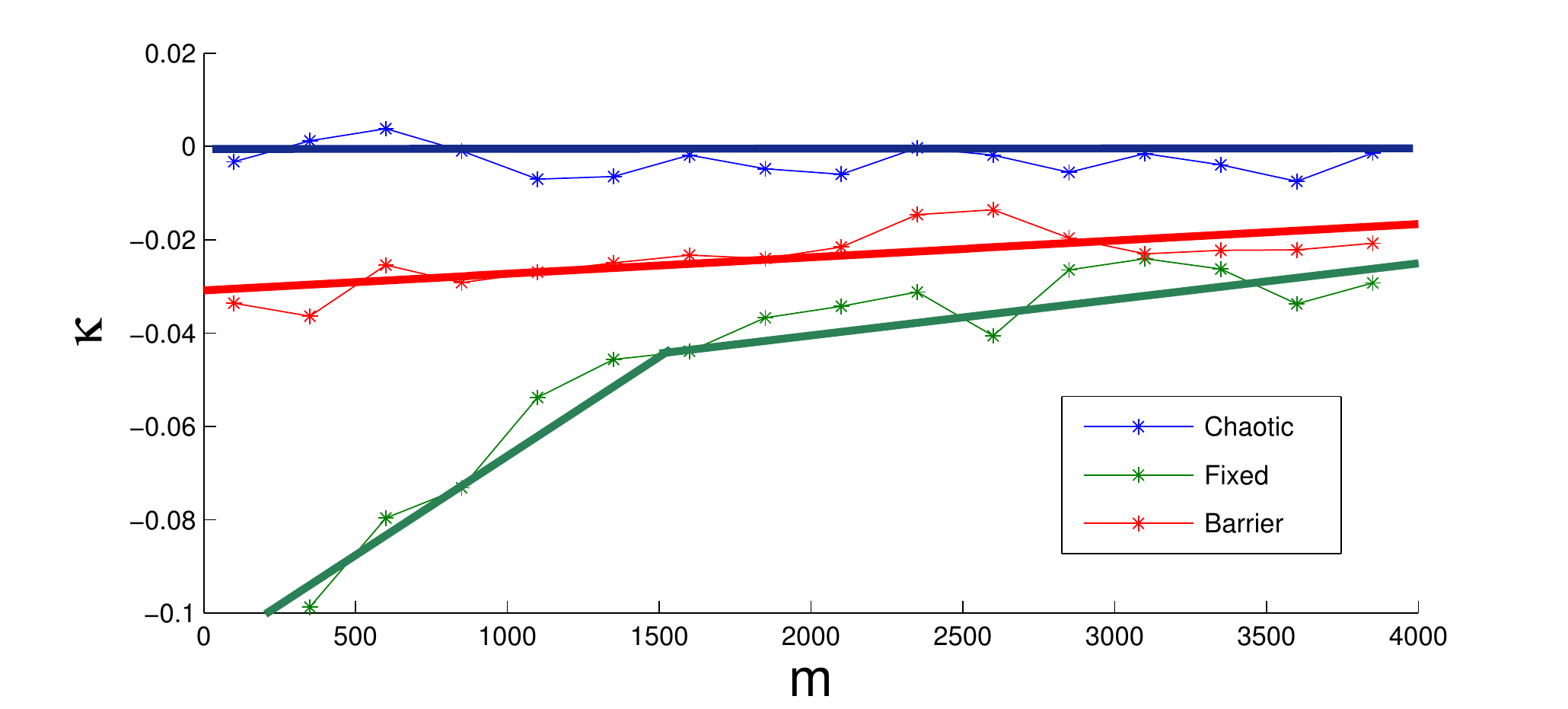} \caption{ $\kappa$ for the observable $g_{1}$ at different bin lengths $m$
for three different ensemble of points. The theoretical expected values
is $\kappa=0$ (black line). See text for a description.}

\label{points} 
\end{figure}

This qualitative considerations can be made quantitative when the
bin length is varied.  This way, one can reliably compute the time scale on which the
barriers become mixing. In order to do so, we consider three different
ensembles of 500 points each. The first one contains all points extracted
in the periodic sea, the second one includes only points located on
the barriers and the third one points in a neighborhood of the fixed
point. For each ensemble, we computed the average shape parameter $\kappa$
at several bin lengths. The results are displayed in Fig.~\ref{points}
together with a linear fit of the data. For the chaotic points, no
substantial dependence on the $m$ chosen is visible and the value
are substantially distributed around the expected value $\kappa=0$.
The points located in the barriers show divergent values of the parameters and,
  for increasing $m$,   approach a fully chaotic behavior  
  extrapolated at $m=10000$. The shape parameter is highly
divergent in the neighborhood of the fixed points and two different
linear approximation have been computed. Note that, if exactly the
fixed point is considered, the fit does not improve even at higher
$m$ but oscillates on negative values. The explanation for the direction
of the drift (towards more negative values of $\kappa$) follows
the argument described in \citet{farandamanneville}.

\section{Final Remarks}

In this paper we have defined a rigorous approach for the definition of mixing time scales by exploiting the results of the extreme value theory for dynamical systems. With respect to classical methods based on the Poincar\'e recurrences theory, our method is able to discriminate between slow mixing regions - the barriers - and the fixed points of the dynamics. Previous analysis (see  \citet{bachelard2007targeted} ) could not highlight any difference between fixed points and barriers by using the statistics of Poincar\'e recurrences. The effectiveness of our method is based on the fact that the asymptotic statistics can be computed analytically just by knowing the geometrical properties of the system (the local dimensions). Once the asymptotic parameters are known, a mixing time is intuitively defined as the minimum bin length of the block-maxima approach such that one obtains convergence to the predicted extreme value laws. Differences between barriers and fixed points appear because asymptotic laws are strongly modified in the proximity of unstable fixed points by the existence of a regular dynamics which is responsible for the clustering of extreme events. Clusters introduce an extra parameter in the theory, the so-called extremal index, which we used indirectly for discriminating between fixed points and barriers.\\

The  results obtained in this paper can be extended to a large class of systems where the computation of mixing time scales are of any interest. Moreover, one has a powerful tool to study the dynamics around unstable fixed points. In a future publication we will address the issue of having an extremal index different than one. In particular, this implies an interesting non-equivalence between the block-maxima approach discussed in the present paper and the peak-over-threshold approach.\\
Other extensions of our methods concern the applicability on   geophysical flows. An example is given in \cite{faranda2013recurrence}. There, mixing-time scales are linked to the definition of normal  or extreme recurrences of air temperature data.  It will be interesting to apply the findings of this paper to extend the results presented in \cite{faranda2013recurrence} and, by including other atmospheric variables, construct a more complex geography of the phase space.

\section{Acknowledgments}

SV was supported by the ANR- Project {\em Perturbations}, by the
PICS ( Projet International de Coop\'eration Scientifique), Propri\'et\'es
statistiques des syst\'emes dynamiques d\'eterministes et al\'eatoires,
with the University of Houston, n. PICS05968 and by the projet MODE
TER COM supported by Region PACA, France. SV and DF acknowledge the
Newton Institute in Cambridge where this work was completed during
the program {\em Mathematics for the Fluid Earth}. DF acknowledges the support of a CNRS post-doctoral grant.

\bibliographystyle{apsrev4-1}
\bibliography{barrier}

\end{document}